\documentclass[pre,
               showpacs,
               superscriptaddress,
               longbibliography
               ]{revtex4-1}

\usepackage{graphicx}
\usepackage{dcolumn}
\usepackage{bm}
\usepackage{amsmath}
\usepackage{amssymb}
\usepackage{color}

\begin{document}

\title{Life stages of wall-bounded decay of Taylor-Couette turbulence}

\author{Rodolfo Ostilla-M\'{o}nico}\email{rostillamonico@g.harvard.edu}
\affiliation{School of Engineering and Applied Sciences, Harvard University, Cambridge, MA 02138, USA.}
             
\author{Xiaojue Zhu}
\affiliation{Physics of Fluids Group, Faculty of Science and Technology, MESA+ Research
             Institute, and J. M. Burgers Centre for Fluid Dynamics,
             University of Twente, PO Box 217, 7500 AE Enschede, The Netherlands.}

\author{Vamsi Spandan}
\affiliation{Physics of Fluids Group, Faculty of Science and Technology, MESA+ Research
             Institute, and J. M. Burgers Centre for Fluid Dynamics,
             University of Twente, PO Box 217, 7500 AE Enschede, The Netherlands.}

\author{Roberto Verzicco}
\affiliation{Dipartimento di Ingegneria Industriale, University of Rome ``Tor Vergata", Via del Politecnico 1, Roma 00133, Italy}
\affiliation{Physics of Fluids Group, Faculty of Science and Technology, MESA+ Research
             Institute, and J. M. Burgers Centre for Fluid Dynamics,
             University of Twente, PO Box 217, 7500 AE Enschede, The Netherlands.}

\author{Detlef Lohse}
\affiliation{Physics of Fluids Group, Faculty of Science and Technology, MESA+ Research
             Institute, and J. M. Burgers Centre for Fluid Dynamics,
             University of Twente, PO Box 217, 7500 AE Enschede, The Netherlands.}
\affiliation{Max Planck Institute for Dynamics and Self-Organisation, 37077 G\"ottingen, Germany. }

\date{\today}

\begin{abstract}

The decay of Taylor-Couette turbulence, i.e~the flow between two coaxial and independently rotating cylinders, is numerically studied by instantaneously stopping the forcing from an initially statistically stationary flow field at a Reynolds number of $Re=3.5\times 10^4$. The effect of wall-friction is analysed by comparing three separate cases, in which the cylinders are either suddenly made no-slip or stress-free. Different life stages are observed during the decay. In the first stage, the decay is dominated by large-scale rolls. Counterintuitively, when these rolls fade away, if the flow inertia is small a redistribution of energy occurs, the energy of the azimuthal velocity behaves non-monotonically: first decreasing by almost two orders of magnitude, and then increasing during the redistribution. The second stage is dominated by non-normal transient growth of perturbations in the axial (spanwise) direction. Once this mechanism is exhausted, the flow enters the final life stage, viscous decay, which is dominated by wall-friction. We show that this stage can be modeled by a one-dimensional heat equation, and that self-similar velocity profiles collapse onto the theoretical solution.
\end{abstract}

\pacs{47.27.nf, 47.32.Ef}

\maketitle

Turbulence is a classic example of a non-equilibrium phenomenon: it requires a constant energy injection as energy is constantly dissipated by viscous effects. For statistically stationary turbulence, the classical picture by Richardson \cite{ric20} and Kolmogorov \cite{kol41a,kol41b,kol41c} is that energy is injected at the larger scales, cascades down to smaller scales and is dissipated by viscosity. In this picture, only the large scales are dependent on the boundary conditions, the forcing and the geometry of the flow. The intermediate and small scales are assumed to be homogeneous and isotropic, and thus the energy dissipation mechanisms are postulated to be universal and self-similar.

Once the forcing is turned off, turbulence decays as energy is no longer injected into the system, but it is still dissipated. Decaying turbulence has been employed not only to study unforced systems, but also the mechanisms for energy dissipation in statistically stationary turbulence, including the dissipation anomaly, a cornerstone of every theory of turbulence \cite{vas15}. Historically, studies have focused on the decay of statistically homogeneous isotropic turbulence (HIT), performed both experimentally (using grid-induced turbulence) \cite{smi93a,sta99} or numerically (using a triply periodic box forced randomly) \cite{geo09}. Some theoretical power-laws for the decay of vorticity and kinetic energy can be obtained both from dimensional analysis, or from a Navier-Stokes equation-based mean field theory \cite{loh94a}, and are in good agreement with available data \cite{smi93a,sta99,ll87}. However, the dependence of decay on initial conditions \cite{lav07,val07,hur11,tei09}, and the degree of self-similarity of the decay have been debated in the literature \cite{saf67,geo09,bif03,eyi00}.

Real systems however are very far from homogeneous and isotropic. Most flows in technology and industry are wall-bounded and thus anisotropic. A considerable part of the energy dissipation occurs next to the walls. And even in geo- and astrophysical flows, where walls might be absent, strong anisotropies still exist as in accretion disks or mantle convection. 

Studies of turbulence decay outside HIT are more limited \cite{tou02}. Taylor-Couette (TC) flow, i.e.~the turbulent flow between two concentric and independently rotating cylinders \cite{gro16}, is an ideal system to study the decay of wall-bounded turbulence because it is a closed system and the cylindrical geometry allows for experiments with relatively small end-effects and more modest sizes unlike channel or pipe flow. While TC had already been employed to study decay of puffs at low Reynolds numbers \cite{bor10}, or turbulence decay due to linear stabilization \cite{ost14b}, the experiment by Verschoof \emph{et al.} \cite{ver16} focused on the fully unstable turbulent regime, at Reynolds numbers of $Re\sim\mathcal{O}(10^6)$. The cylinders were stopped, and the turbulence was allowed to decay. The decay did not follow a pure power law like HIT, but instead decayed faster due to the viscous drag applied by the walls. Nonetheless, the decay was found to be self-similar. However, the considerable inertia of the massive metal cylinders causes the braking to take a substantial amount of time (approximately 3000 eddy turnover times), during which the decay could not be measured. The experiment also has secondary flows due to residual cooling effects, which become important at later times and cause deviations from the models in Ref.~\cite{ver16}.

In this manuscript, we numerically simulate Taylor-Couette turbulence, analogous to the experiments of Ref.~\cite{ver16}, but now we \emph{instantaneously} stop the cylinders (which is of course possible in numerical simulations) to better understand the initial stages of the decay, and run the simulation until the fully viscous final stage dominates the system dynamics. In addition, we can remove the effect of wall-friction by making the walls stress-free, and thus we can clearly separate the role of the walls from other decay mechanisms in the flow. We find that the decay regime observed and discussed in Ref.~\cite{ver16} corresponds to an intermediate stage of decay. We do not find the proposed self-similar regime until the last life-stage which can be described as a quenching problem. Prior to this last viscosity-dominated stage, two earlier life stages of decay are observed, which are dominated by linear instabilities and non-normal transient growth, respectively.

We simulate the incompressible Navier-Stokes equations using a second-order energy-conserving centered finite difference scheme, which treats the viscous terms semi-implicitly \cite{ver96,poe15}. The initial starting field corresponds to a TC system driven by pure inner cylinder rotation in an inertial reference frame. To reduce dispersive errors in the simulations, we simulate the system in a rotating frame such that the velocities at both cylinders are always equal and of opposite signs ($\pm U/2$). Due to this reference frame change, a Coriolis force naturally arises \cite{ost14d}. The initial shear Reynolds number was taken as $Re_s = Ud/\nu = 3.52\times 10^4$, where $r_i$ ($r_o$) is the inner (outer) cylinder radius, $d$ the gap width $d=r_o-r_i$, and $\nu$ the kinematic viscosity of the fluid.  Periodic boundary conditions are employed in the axial direction. The system has a radius ratio of $\eta=r_i/r_o=0.909$ and an aspect ratio of $\Gamma=L_z/(r_o-r_i)=4$, where $L_z$ is the axial periodicity length. This geometry was chosen such that the system had two fixed roll pairs (i.e.~four rolls) and curvature effects are limited.  A rotational (azimuthal) symmetry of order $n_{sym}=10$ was forced on the system to reduce computational costs. This means that the system is $4\pi$ half-gaps long in the streamwise direction at its smallest extent. A resolution of $N_\theta \times N_r \times N_z = 1024\times512\times1024$ was used, the grid nodes were uniformly distributed in the axial and azimuthal directions, while a clipped Chebychev distribution was used for the radial direction to cluster points near the walls. This is equivalent to the resolutions used for previous studies \cite{ost16vlb}. 

The simulations were run until a statistically stationary state was reached, where two roll pairs could be seen in the velocity fields, and the torque at both cylinders was equal when temporally averaging over a sufficiently long time.  We simulate three cases: for the first case, which we denote as sudden stop as in the experiment (SS-EXP), we instantaneously stop the inner cylinder at $t=0$, and allow the turbulence to decay. The main difference between the simulation and the experiment is that we stop the cylinders instantaneously, while the stopping time of the experiment by Verschoof \emph{et al.}~\cite{ver16} is around 12 s. Even if this is still orders of magnitude smaller than the viscous time ($d^2/\nu=6\cdot 10^3$ s) it corresponds to over three-thousand large eddy turn-over times as mentioned previously ($d/U=3.2\cdot10^{-3}$ s). 

To eliminate as much as possible the effects of the wall, we run a second case, denoted by sudden disengagement (SD) in which we suddenly change the boundary condition at the wall to stress-free. The flow is no longer forced by shear, and thus the turbulence decays. Finally, we run an intermediate, third case, denoted by (SS-V0) where we set the velocity of both cylinders to the mean velocity, i.e. zero in the rotating frame. While in the experiment, and in the SS-EXP simulation, the inner cylinder is brought to the same velocity as the outer cylinder, and there is a substantial mean momentum which must be transported from bulk to walls, this is not the case in the SS-V0 case because the mean momentum respect to the cylinders is close to zero. In both the SS-V0 case and the SD case, the main decay is that of velocity \emph{fluctuations} while in the SS-EXP case it is of the \emph{mean} azimuthal (streamwise) velocity. Thus, we expect the SS-V0 case to behave in an intermediate way between the SD case and the SS-V0 case.

As turbulence decays, the computational method is switched for efficiency. At $\tilde{t} = tU/d \approx 125$ for the SS-V0 case, and $\tilde{t} \approx 325$ for the SD and SS-EXP cases, the flow fields were down-sampled to a resolution of $512^3$, and the treatment of the viscous terms in the homogeneous directions was made implicit. This allowed the time-step to increase to $\Delta \tilde{t} = 1$, dramatically reducing the computational cost. The simulations were then advanced in time up to $\tilde{t}=7500$ (SS-V0/SD) and $\tilde{t}=15000$ (SS-EXP), late enough to be in the asymptotic viscous stage.

\begin{figure}
\includegraphics[width=0.48\textwidth]{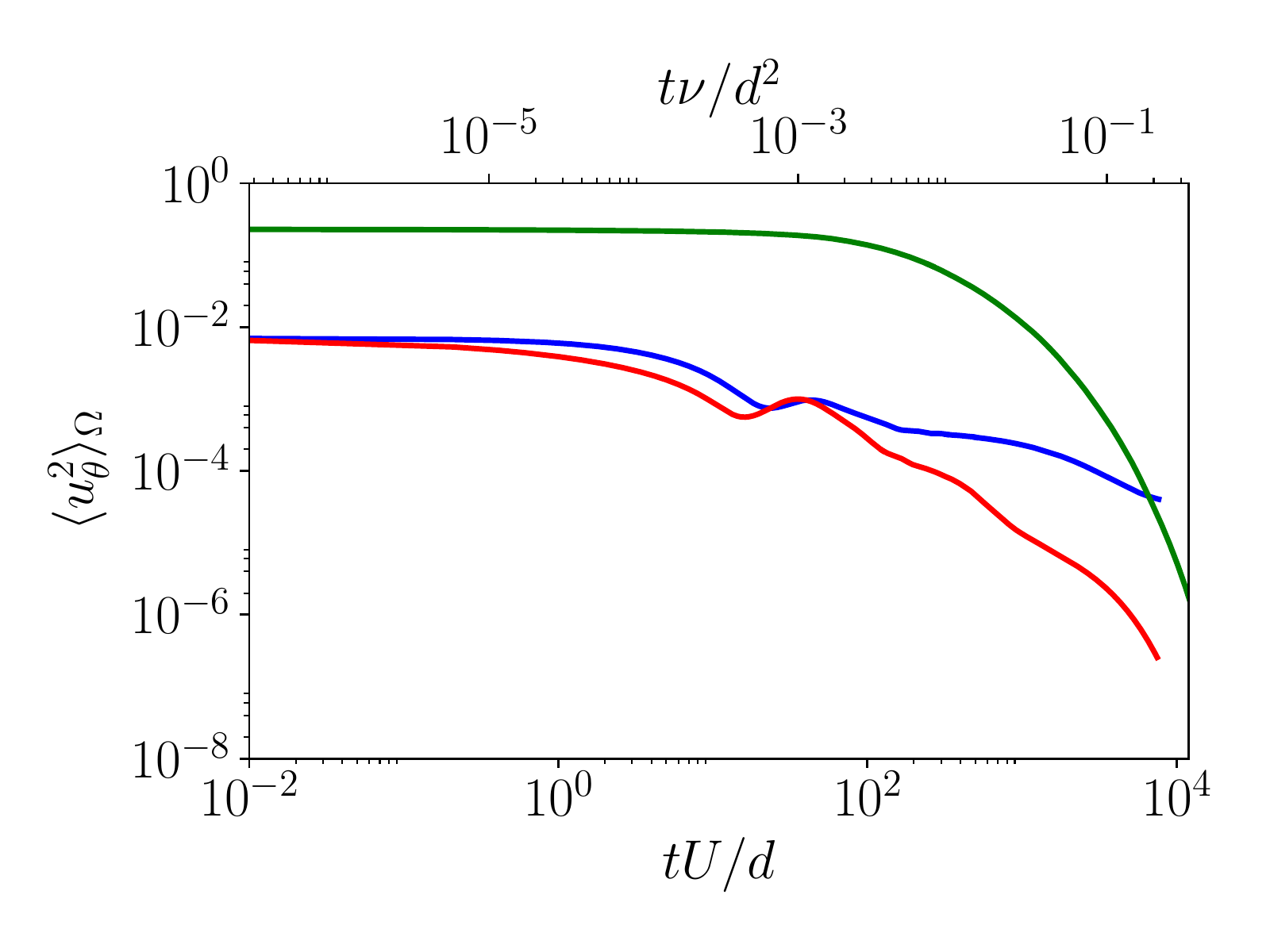}
\includegraphics[width=0.48\textwidth]{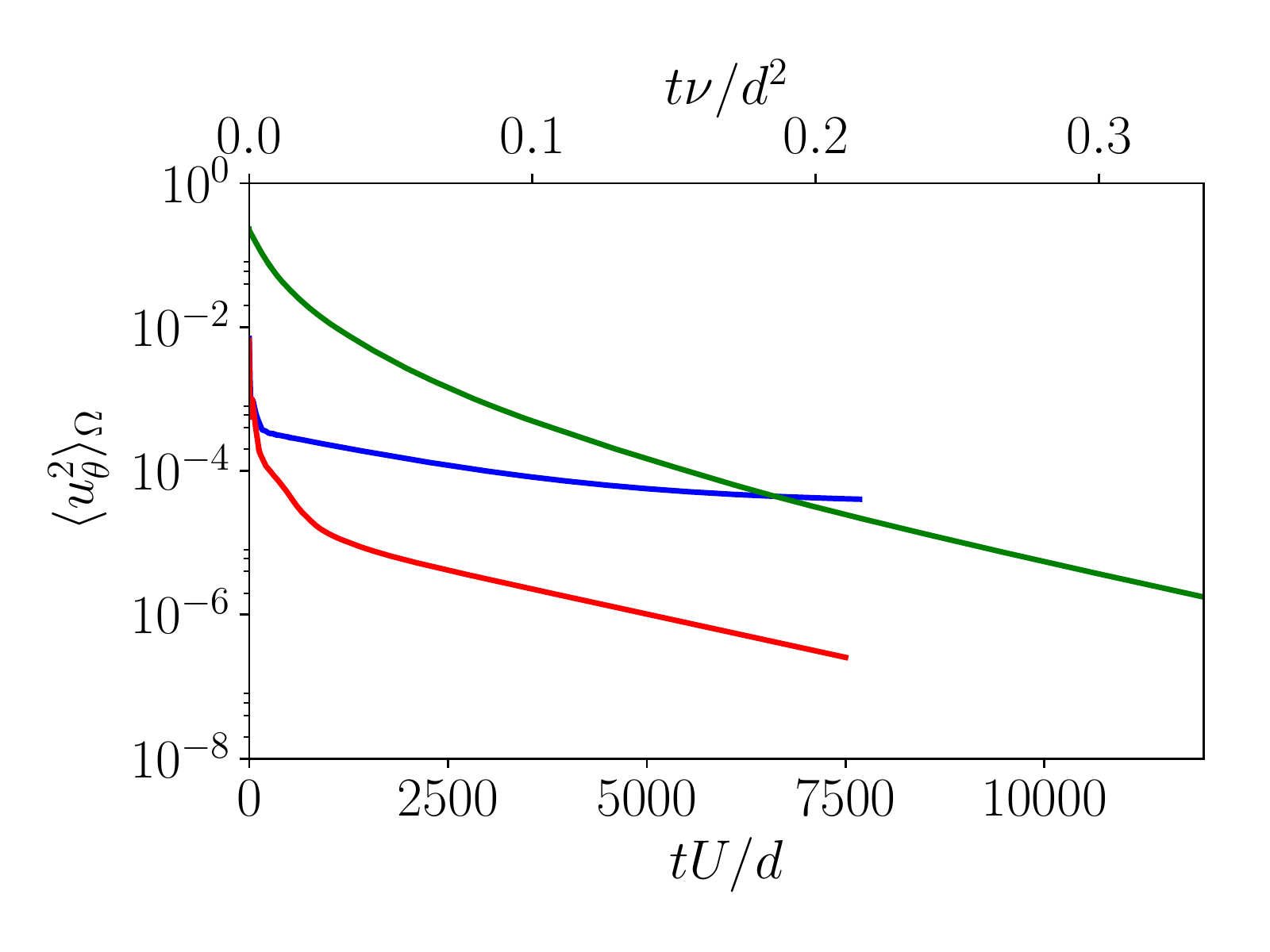}\\
\includegraphics[width=0.48\textwidth]{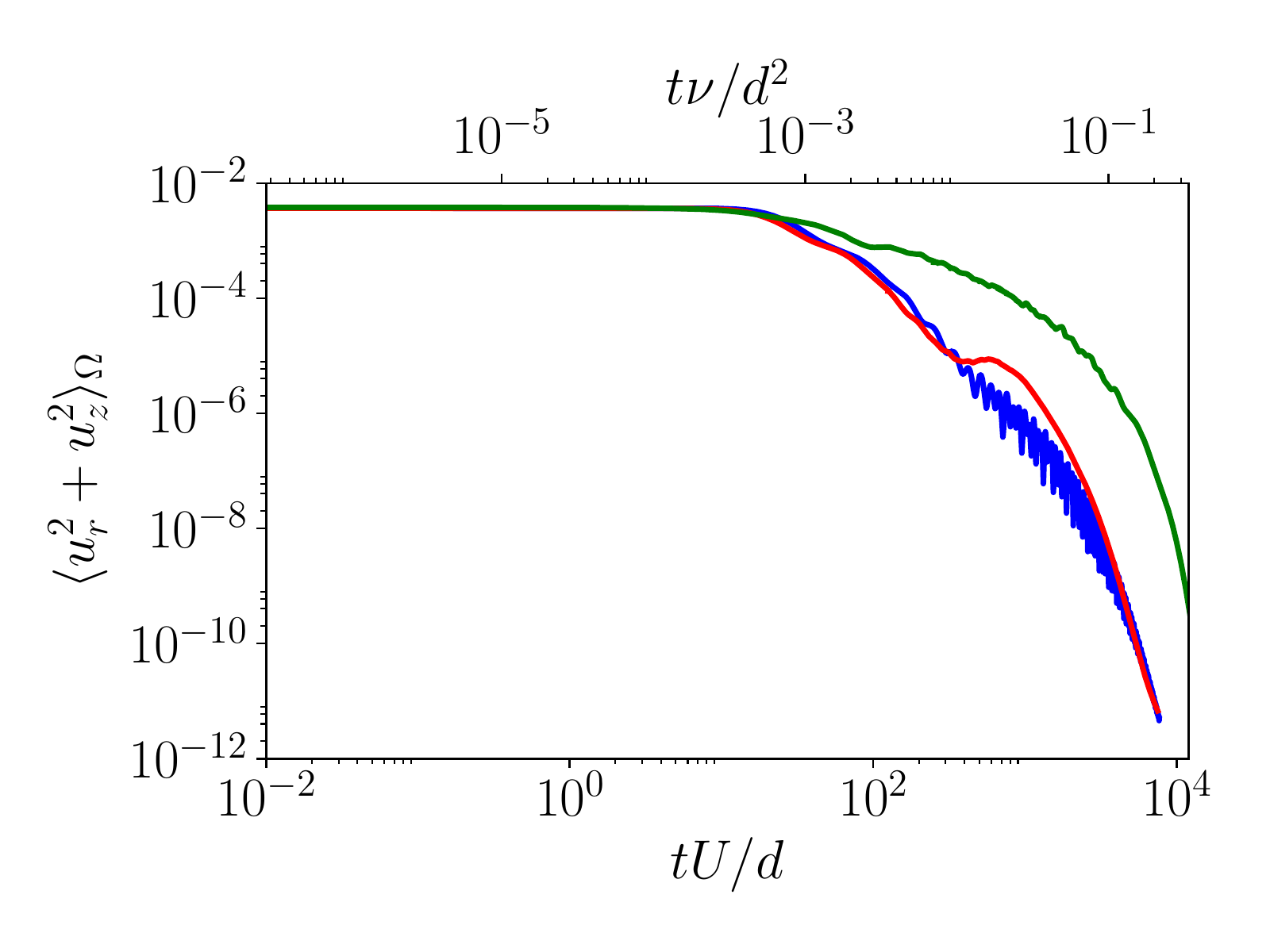}
\includegraphics[width=0.48\textwidth]{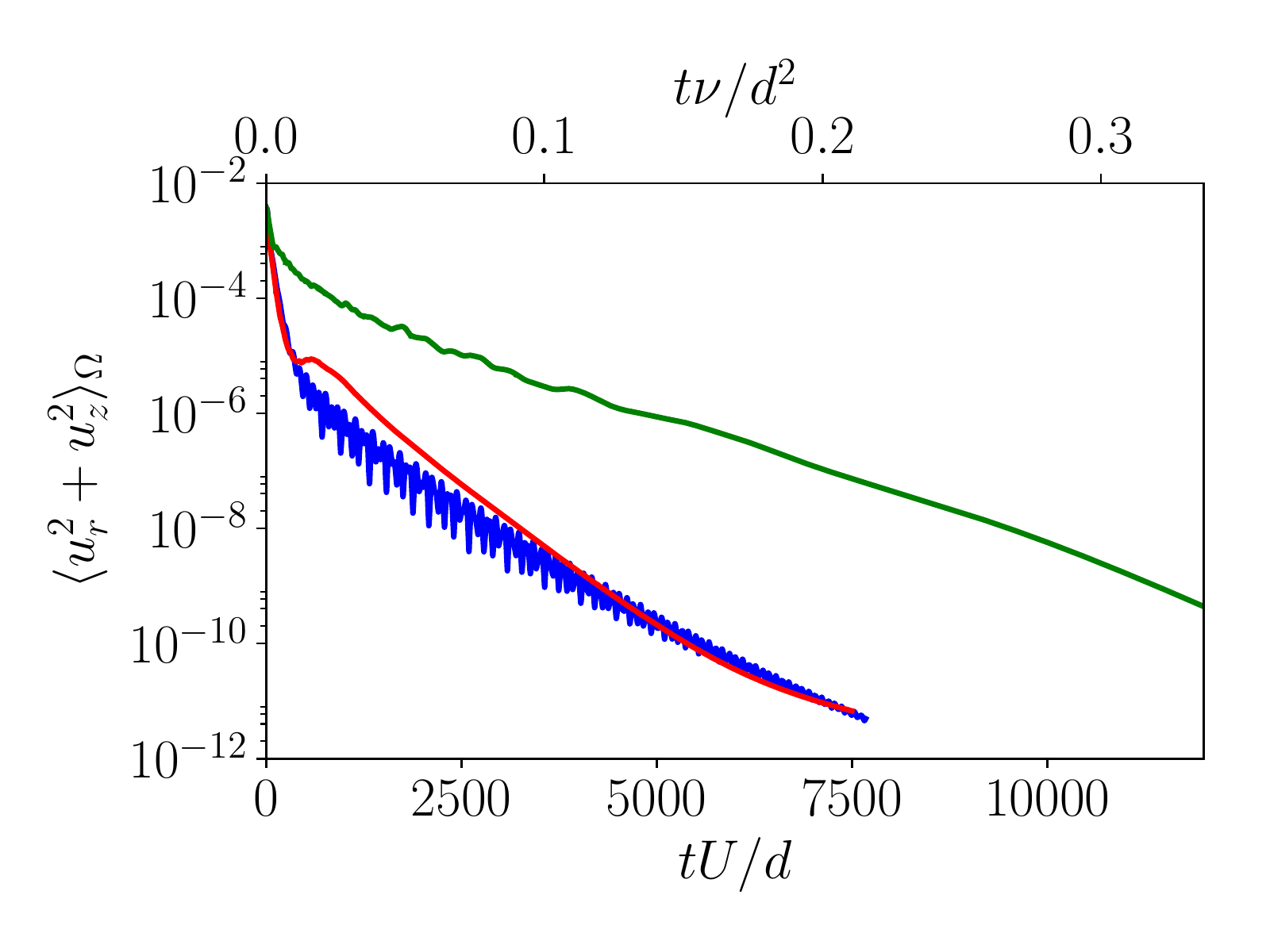}\\
\centering
\caption{Top left: Temporal evolution of the kinetic energy in the azimuthal velocity. Bottom left: Temporal evolution of the wind kinetic energy. The figures show the same data in semilogarithmic scale (right) and double logarithmic scale (left). Symbols: SS-EXP (green), SD (red) and SS-V0 (blue). }
\label{fi:urms} 
\end{figure}

\begin{figure}
\includegraphics[width=0.48\textwidth]{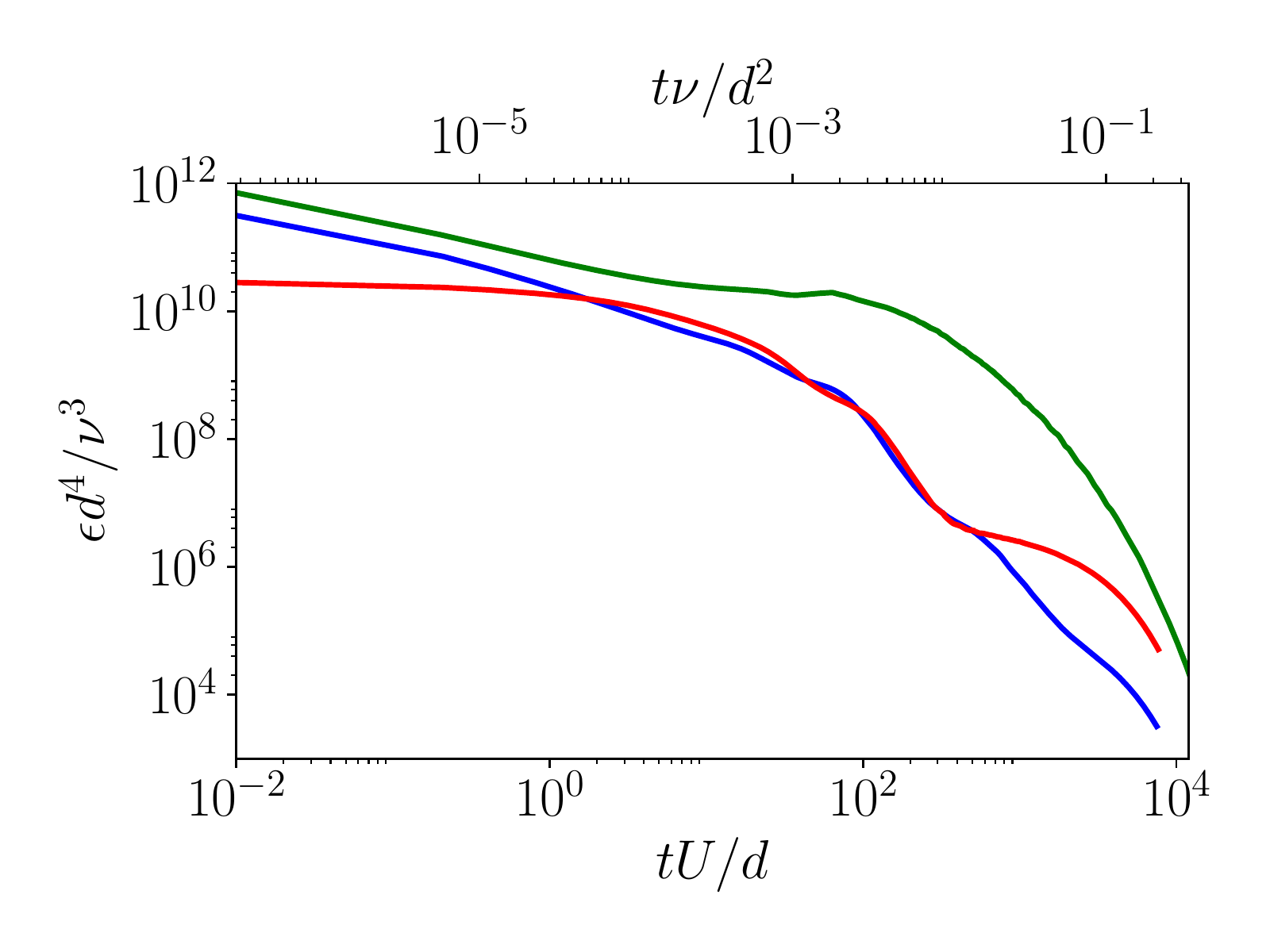}
\includegraphics[width=0.48\textwidth]{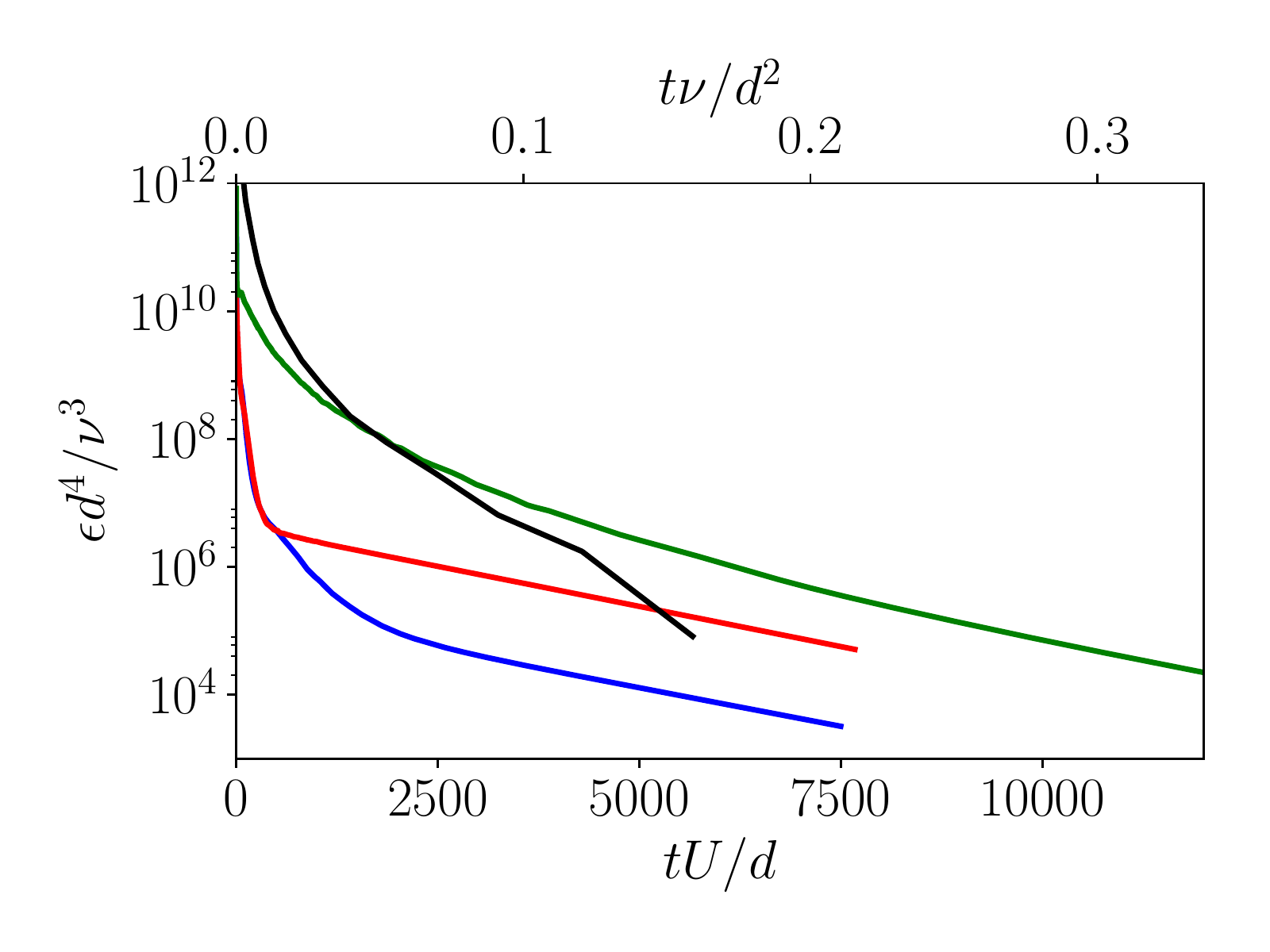}\\
\centering
\caption{Temporal evolution of the energy dissipation. The figures show the same data in semilogarithmic scale (right) and double logarithmic scale (left). Symbols: SS-EXP (green), SD (red), SS-V0 (blue), experimental data from Ref.~\cite{ver16} (right, data matches top temporal axis only). }
\label{fi:diss} 
\end{figure}

\begin{figure*}
\centering
\includegraphics[width=0.8\textwidth]{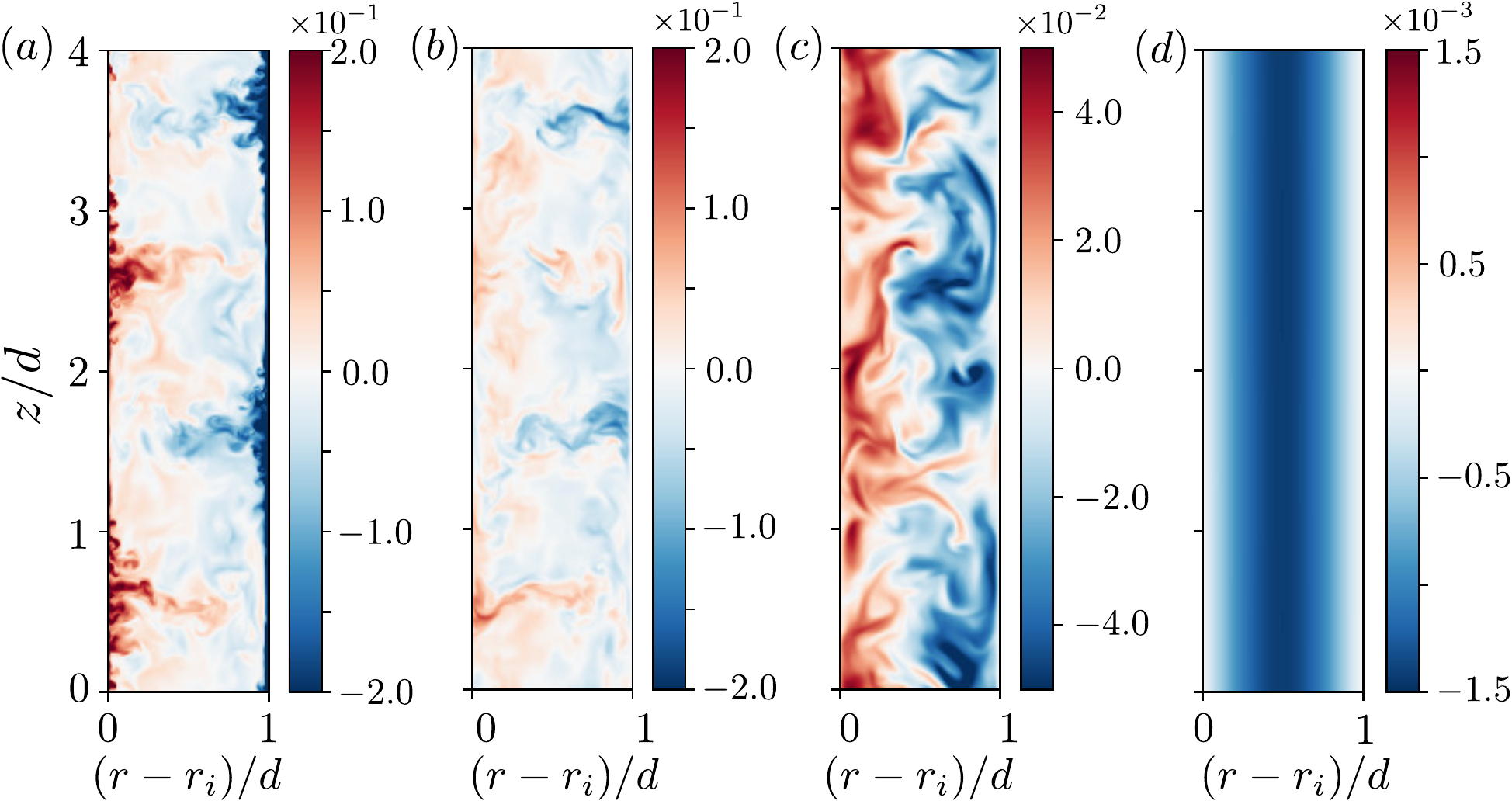}
\caption{Pseudocolor visualization of the azimuthal velocity in a constant-azimuth cut for the SS-V0 case at four different times (left to right): $\tilde{t}=0$, $10$, $75$ and $5000$. }
\label{fi:vizthcutnsvzero} 
\end{figure*}

\begin{figure}
\centering
\includegraphics[width=0.98\textwidth]{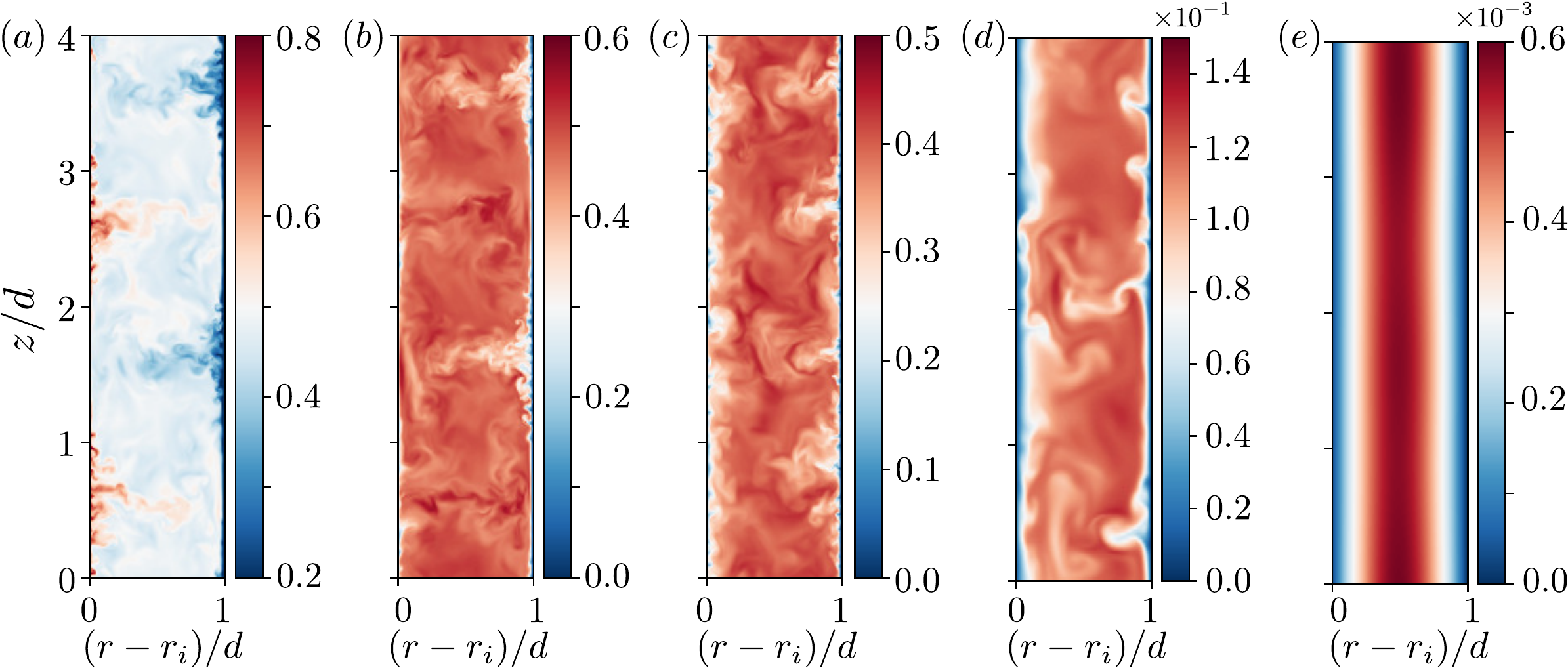}

\caption{Pseudocolor visualization of the azimuthal velocity in a constant-azimuth cut for the SS-EXP case at five different times (left to right): $\tilde{t}=0$, $10$, $100$, $1000$ and $17000$. }
\label{fi:vizthcutnsexp} 
\end{figure}

Figure \ref{fi:urms} shows the temporal evolution of the average kinetic energy of the azimuthal velocity ($\frac{1}{2}\langle u_\theta^2\rangle_\Omega$) and the wind kinetic energy ($\frac{1}{2}\langle u_r^2 + u_z^2 \rangle_\Omega$) for all cases, while figure \ref{fi:diss} shows the temporal evolution of the total kinetic energy dissipation rate $\epsilon=\frac{1}{2}\nu\langle \partial_i u_{j}\rangle^2_\Omega$, where $\langle \dots \rangle$ denotes the averaging operator and $\Omega$ the entire fluid volume. On the right panel of figure \ref{fi:diss}, experimental data from \cite{ver16} are also added for comparison. This data only matches the top temporal axis (viscid units).

No overarching behavior or power law which describes either the dissipation or the evolution of the kinetic energy can be seen. Instead, several different life stages of the decay are revealed, which we detail here. In figure \ref{fi:urms}, the main difference between the SS-EXP and the other two cases can be clearly seen: the azimuthal kinetic energy (in the bulk) is several orders of magnitude larger than in the other cases. Throughout the entire decay, it is significantly larger than the wind kinetic energy, in line with what was observed by Ref.\cite{ver16}. As we will see, this dominates the physics of the decay. The black curve in the right panel of Fig.~\ref{fi:diss} shows a much faster decay in the dissipation rate (when measured in viscous time units) than any of the numerical cases simulated, the reasons for this will be explored later.

The stages of decay are illustrated by Figures \ref{fi:vizthcutnsvzero}-\ref{fi:vizthcutnsexp}, which shows a visualizations of the azimuthal velocity at different time instants for the SS-V0 and SS-EXP cases, from the start of the simulation. Aside from the magnitude of the velocity being different, we can also see wide variation in the flow topology. A similar flow topology as the one seen for SS-V0 is seen for the SD case, too.

We briefly describe the stages for the SS-V0 and SD cases: during the first life stage, which takes place between $\tilde{t}=0$ to $\tilde{t}\approx10$ (corresponding to Figure \ref{fi:vizthcutnsvzero}\emph{b}), the large-scale rolls remain in motion, as they are still being fed from the perturbations inside the boundary layers. After this, the flow undergoes non-linear non-normal transient growth, as in an ordinary shear flow. Finally, in the last, asymptotic stage, (shown in the rightmost panels), viscous diffusion dominates and the flow becomes homogeneous in the azimuthal and axial directions.

The SS-EXP case shows different behaviour, as the large scale roll remains being fed throughout the entire simulation by the outer cylinder, while the inner cylinder undergoes a rapid change between feeding the roll through a centrifugal instability for $\tilde{t}<0$, to being centrifugally stable for $\tilde{t}>0$. This is especially noticeable in the second panel of figure \ref{fi:vizthcutnsexp}, where remnants of ejection zones of the outer cylinder are clearly visible, but the flow topology at the inner cylinder has changed. After this phase, the flow loses the clear axial signature of the rolls (third panel), but the strong boundary layer asymmetry remains in place as evident in the fourth panel.  At even later times, shown in the last panel, the flow also enters the asymptotic stage, where it becomes practically homogeneous in the azimuthal and axial direction, analogous to the right panel of figure \ref{fi:vizthcutnsvzero}. 

No significant axial signature was observed in Ref.~\cite{ver16} after the long stopping time of the cylinders, which provides some indication that their measurements start at a time where the flow approximately behaves like in the third panel of figure \ref{fi:vizthcutnsexp}.

Due to the less clean delimitation between stages in the SS-EXP case, we first focus on the three stages of the SS-V0 and SD cases. During the first life stage, which takes place between $\tilde{t}=0$ to $\tilde{t}\approx10$ (corresponding to Figure \ref{fi:vizthcutnsvzero}\emph{b}), the large-scale rolls remain in motion, as they are still fed from the perturbations inside the boundary layers. Thhe wind kinetic energy remains constant (horizontal) between $\tilde{t}=0$ and $\tilde{t}\approx 10$, while the kinetic energy of the azimuthal velocity decreases by more than an order of magnitude. The different behavior of velocity components highlights the anisotropic character of the decay. 

We note that the time-scale of this first stage is the same for both the SS-V0 and the SD case, highlighting that the rolls drain energy much more efficiently than the wall. A simple scaling estimate for the diffusive time across the boundary layer of the available kinetic energy to the wall is $t_{\nu,BL} = \lambda_\omega^2/\nu$. If we estimate the boundary layer thickness $\lambda_\omega$ as $d/(2Nu_\omega)$, with $Nu_\omega$, the generalized Nusselt number from the initial turbulent simulation and equal to $Nu_\omega \approx 25$ \cite{ost16vlb}, we obtain an estimate for $t_{\nu,BL} U/d \approx 200$, an order of magnitude larger than the time-scale seen. This stage is not seen in Fig.~\ref{fi:urms} for the SS-EXP case, because the kinetic energy is dominated by the bulk, and not by the boundary layers. From this we would also not expect this stage to be present in the experiment. 

\begin{figure}
\centering
\includegraphics[width=0.48\textwidth]{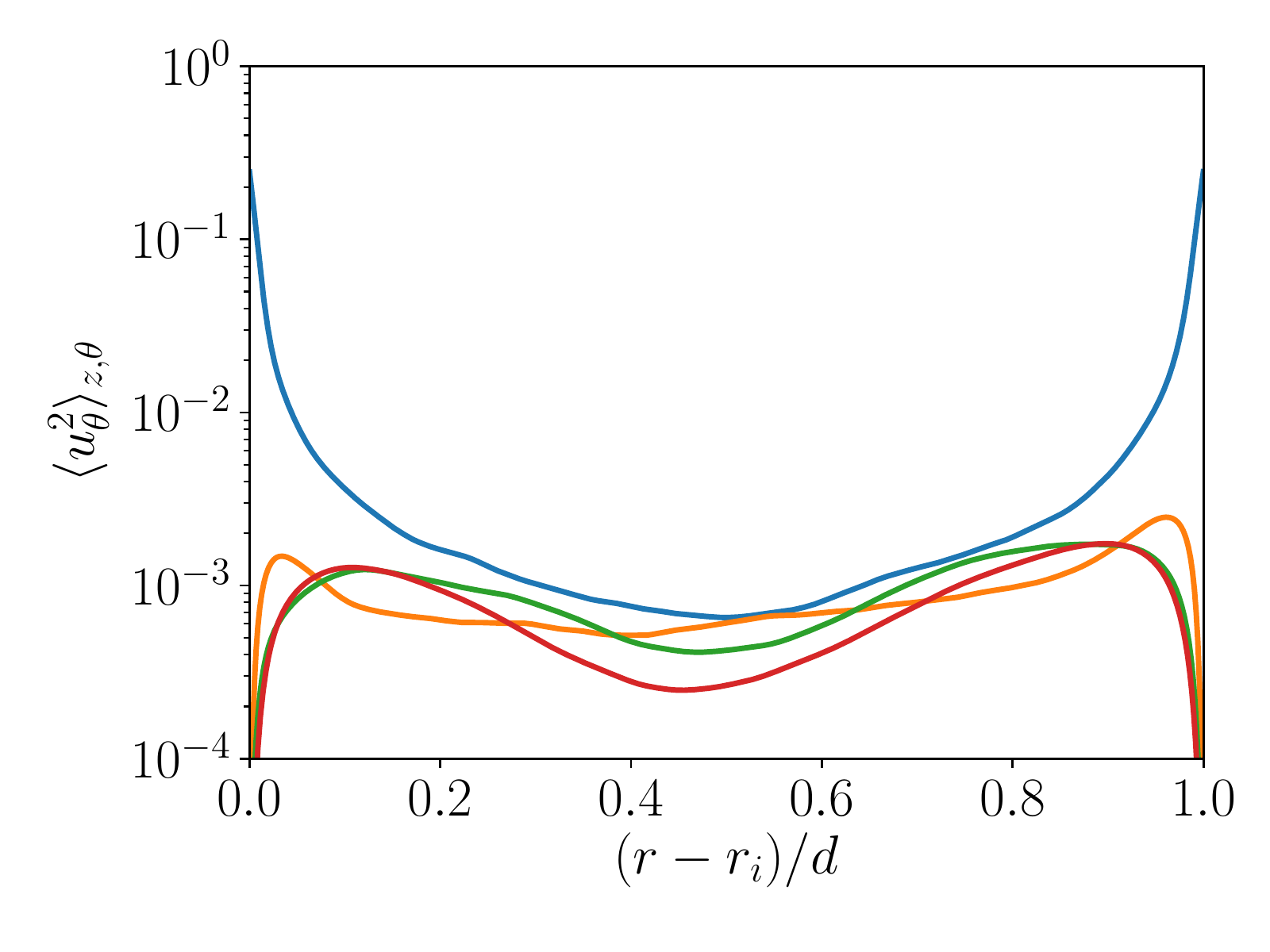}
\includegraphics[width=0.48\textwidth]{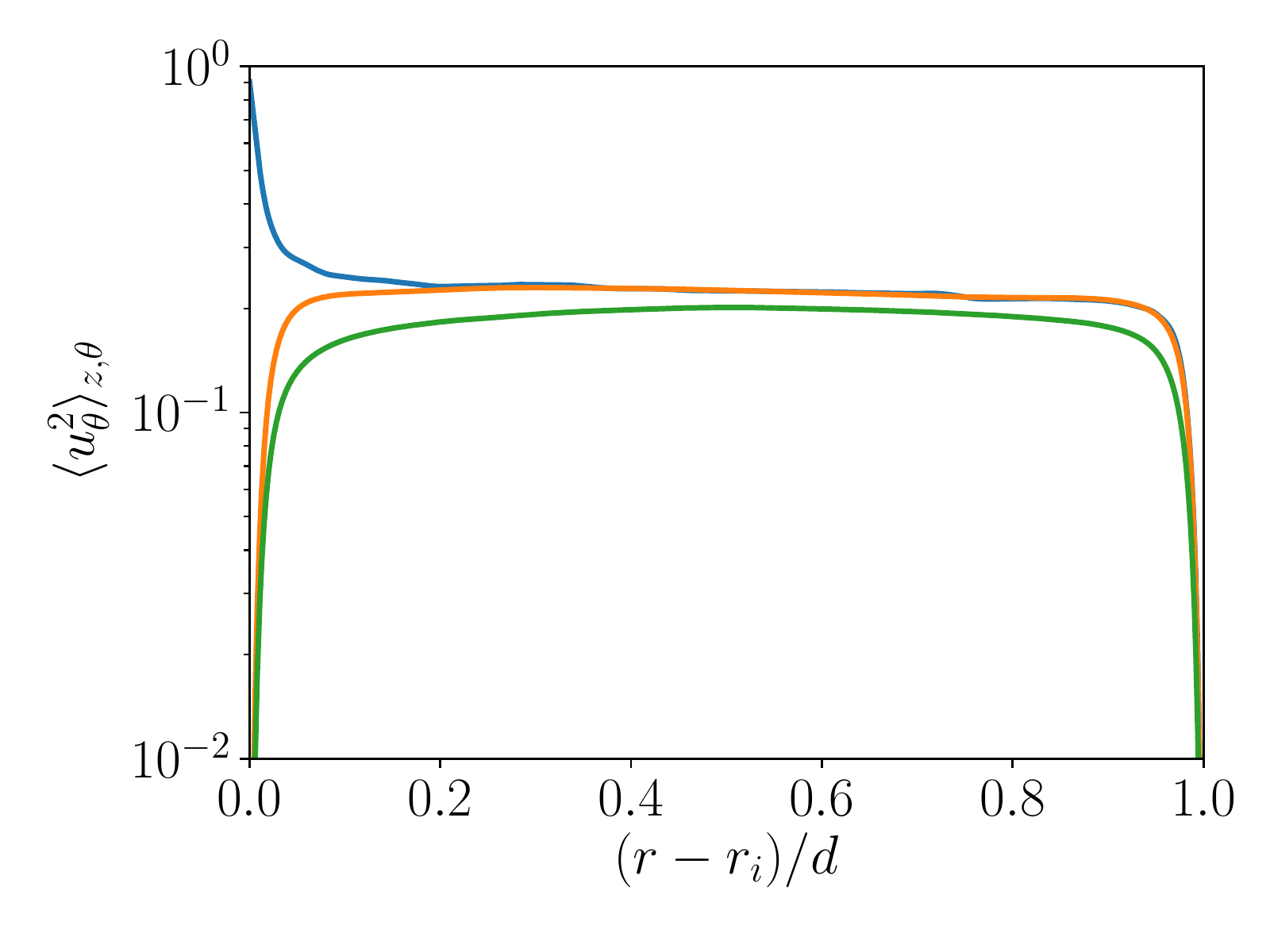}
\caption{Left: Azimuthally and axially averaged azimuthal kinetic energy dissipation for the SS-V0 case, and $\tilde{t}=0$ (blue), $\tilde{t}=10$ (orange), $\tilde{t}=30$ (green) and $\tilde{t}=50$ (red). Right: same, for SS-EXP case and $\tilde{t}=0$ (blue), $\tilde{t}=10$ (orange) and $\tilde{t}=50$ (green) }
\label{fi:kinen_s1} 
\end{figure}

Further proof of this is seen in figure \ref{fi:kinen_s1}, which shows the distribution of the azimuthal kinetic energy at the start and end of the first stage, and at similar times for the SS-EXP case. At the start of the decay, energy is concentrated near the boundary layers. Once this energy is drained by the rolls, they fade away. The boundary layers grow, and the fluctuation maximum moves away from the walls. Energy is redistributed between the existing structures, and the flow topology changes from Figure \ref{fi:vizthcutnsvzero}b to \ref{fi:vizthcutnsvzero}c. Counterintuitively, the decay of azimuthal kinetic energy is non-monotonic, and in Fig.~\ref{fi:kinen_s1}, we see an \emph{increase} in azimuthal kinetic energy at around $\tilde{t}\approx 15$ which is drawn from the wind kinetic energy during the redistribution, up to a maximum at $\tilde{t}\approx 40$. Wall-friction plays a secondary role during this stage. The main player draining the available energy in the boundary layer are the rolls. The wall-friction comes in as a small correction, making the SS case undergo a transition from the first life stage to the second life stage slightly faster than the SD case. Non-monotonic behavior can be seen for both SS and SD cases. For the SS-EXP case, no redistribution is seen. The energy levels for the bulk are much higher and this dominates $\langle u^2_\theta \rangle_\Omega$.

Then the second life stage is entered which takes place between $\tilde{t}\approx 10 $ and $\tilde{t} \approx 500$. With the rolls faded away, perturbations are amplified non-normally. A detailed discussion of this non-normal-non-linear mechanism is found in Ref. \cite{gro00rmp}. Here it is sufficient to state that perturbations in the axial (spanwise) direction tend to grow faster than those in other directions. As typical for this type of instability, the growth of these perturbations is transient, and after several time units they decay. For the SS-V0 case, the wind kinetic energy drops to a minimum around $\langle u^2 \rangle_\Omega \approx 10^{-5}$ and it remains at that level until the end of this regime, while for the SD case, the wind kinetic energy does not reach a plateau but instead oscillations appear throughout the entire decay. The SS case appears to drain the available energy of the many non-normal modes in a shorter amount of time, while this draining occurs over a longer time-scale for the SD case. The signature of non-normal transient decay can be clearly seen for the wind kinetic energy in the SD and SS-EXP cases up to the end of the simulation. When one of the non-normal modes grows, amplifies transiently and decays, it imprints the oscillations seen in the wind kinetic energy for $\tilde{t}>100$.

\begin{figure}
\centering
\includegraphics[width=0.98\textwidth]{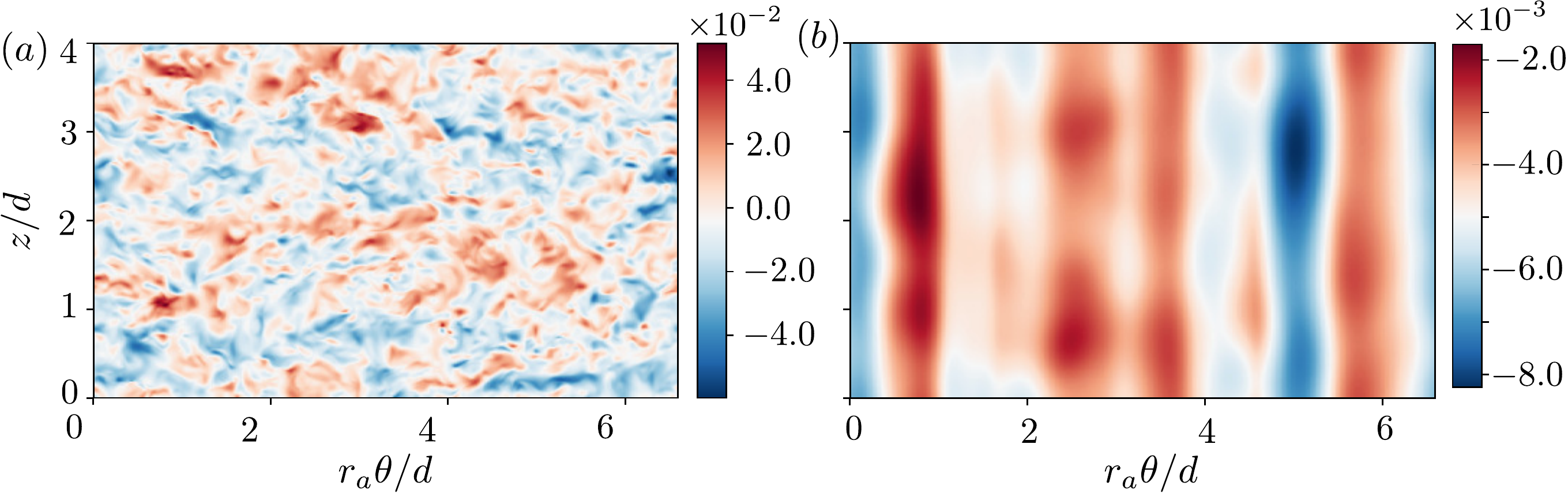}
\caption{Pseudocolor visualization of the azimuthal velocity in a constant-radius cut at the mid-gap ($r=r_a$) for the SS-V0 case at two different times $\tilde{t}=75$ (top) and $1000$ (bottom). }
\label{fi:vizrcut} 
\end{figure}

Due to the non-normal transient growth and dissipation of axially-oriented perturbations, the flow becomes practically homogeneous in the axial direction for $\tilde{t} > 500$ in the SS-V0 case. The large-scale flow structure is now columnar. This is shown in Figure \ref{fi:vizrcut}. The SD and SS-EXP cases do not show this, and instead show both azimuthal and axial dependency even at later times. Unlike the SS-V0 case, the marks of non-linear transient growth seem to be present in the wind kinetic energy even up to $\tilde{t} \approx 2500$ for these two cases. 

The experimental data shows an even steeper decay of $\epsilon$ than the DNS in Fig.~\ref{fi:diss}. The instantaneous Reynolds number is still significantly higher in the experiment than in the simulations even for later times $Re\approx 500$. Significant non-linear non-normal transient growth can still happen, which means an overall faster draining of energy and an overall faster decay rate (when looking at the flow in viscid time units). This is reflected in a steeper decay of $\epsilon$. 

The role of wall-friction is minor in the transition between the second and third stage, as this happens when transient non-normal growth is exhausted. Once either the non-linear transient growth mechanism is drained, or the wind kinetic energy is sufficiently small that its effect cannot be felt, the decay enters its last life stage, in which the decay is dominated by viscosity. 

\begin{figure}
\centering
\includegraphics[width=0.48\textwidth]{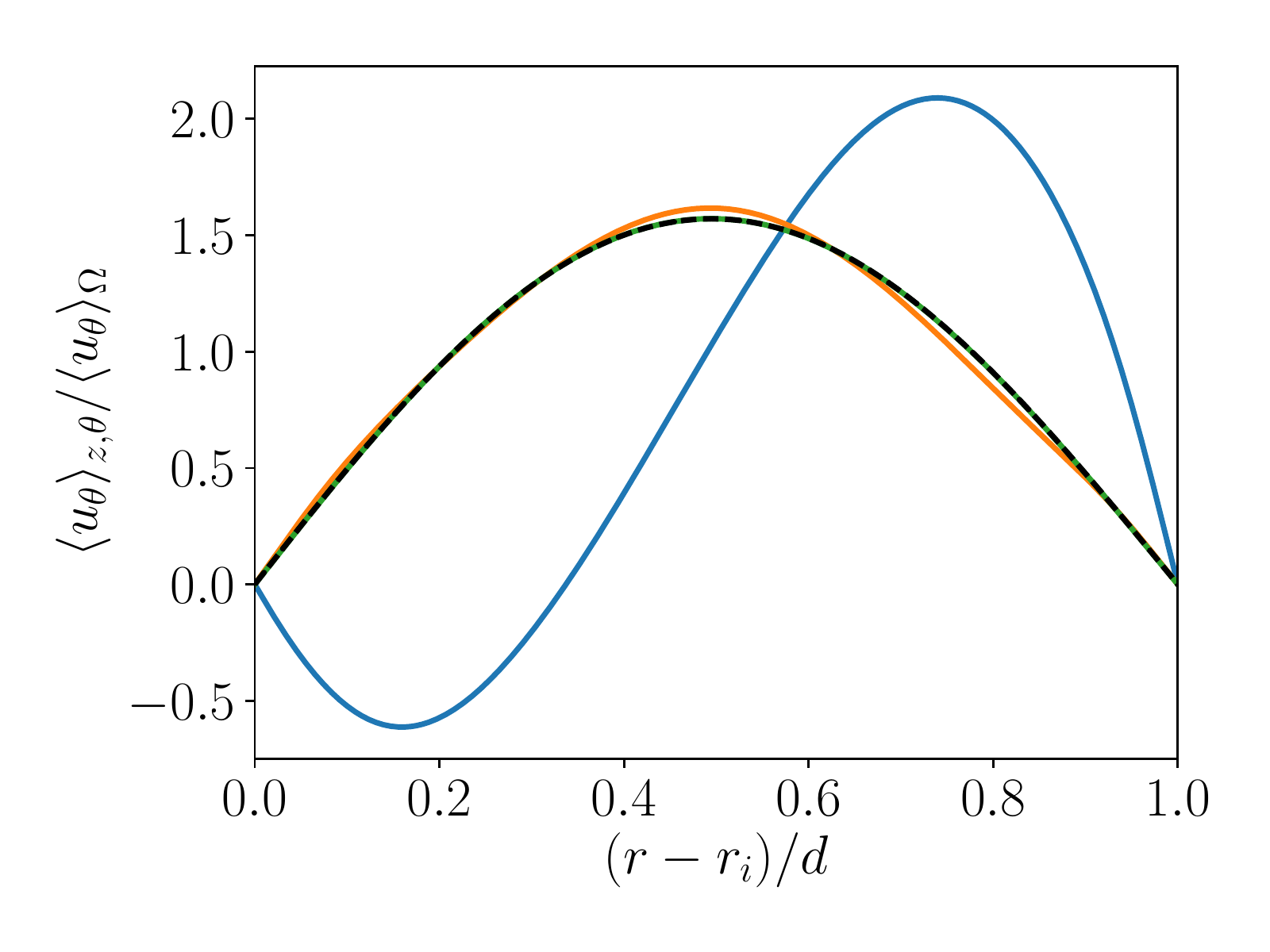}
\includegraphics[width=0.48\textwidth]{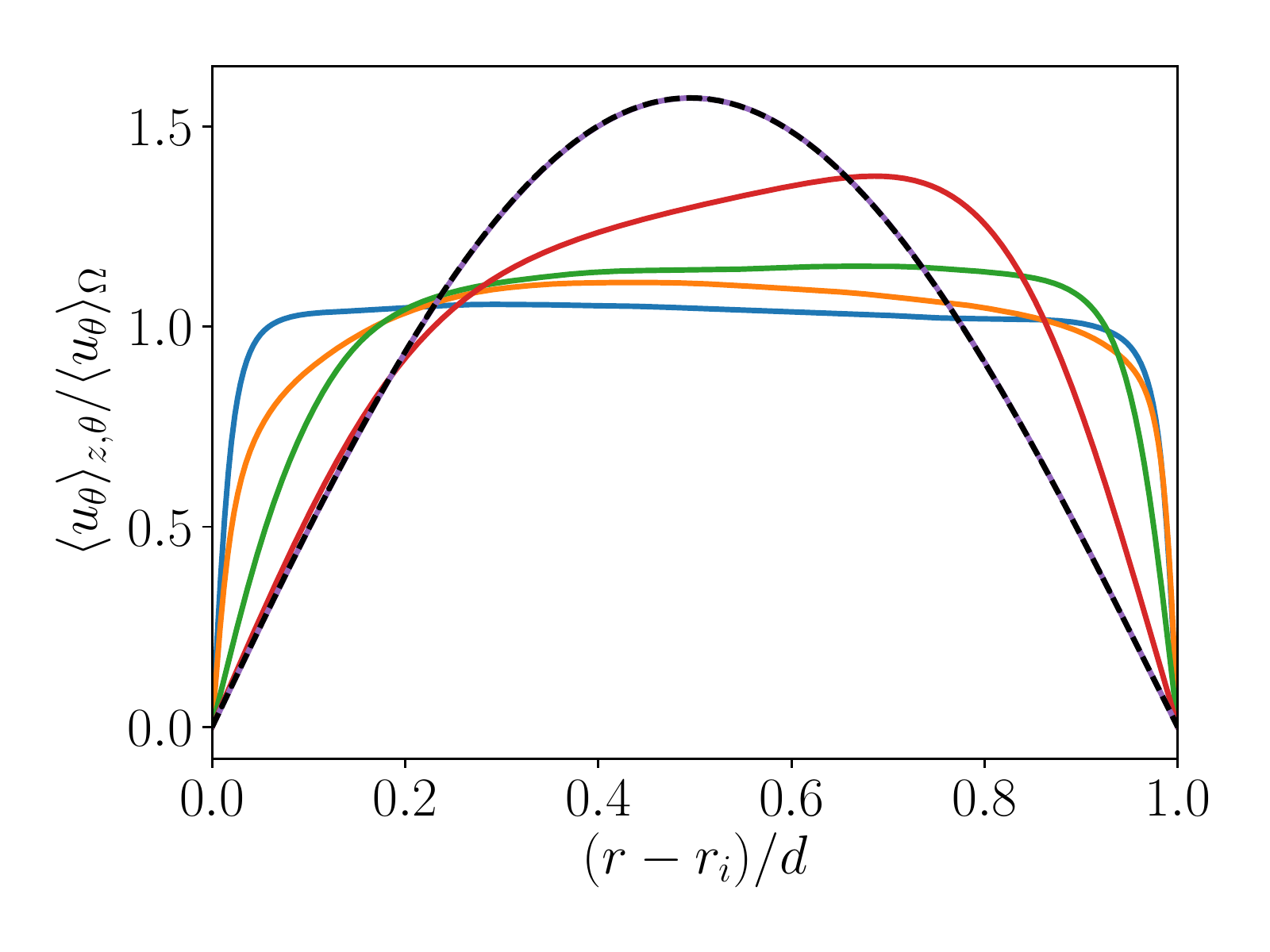}
\caption{Left: Axially and azimuthally averaged normalized azimuthal velocity profiles for the SS-V0 at $\tilde{t}=500$ (blue), $\tilde{t}=1000$ (green) and $\tilde{t}=5000$ (orange), as well as the theoretical solution based on Bessel functions (black dashed). Right: same, for the SS-EXP case at times $\tilde{t}=10$ (blue), $\tilde{t}=100$ (orange), $\tilde{t}=1000$ (green), $\tilde{t}=5000$ (red) and $\tilde{t}=15000$ (purple). }
\label{fi:selfsim} 
\end{figure}

In this last life stage the energy dissipation rate behaves in a quasi-exponential manner. The wind kinetic energy is negligible and we can expect the azimuthal velocity to behave like a passive scalar with diffusion, i.e. a heat equation. With the boundary and initial conditions, the problem is analogous to the quenching of a cylindrical annulus, whose solution is given by $u_\theta(r,t)=\sum_n A_n J_0(\lambda_n r)\exp(-\nu \lambda_n^2 t)$, where $J_0(r)$ are Bessel functions of the zeroth order and $\lambda_n$ their associated eigenvalues, and $A_n$ depend on the initial conditions \cite{prosbook}. If $t$ is large, only the slowest-decaying function with the smallest eigenvalue has significant energy. The energy dissipation behaves exponentially in time, and self-similarity comes out naturally. 

To demonstrate that for very large times (beyond $\tilde{t} > 5000$ for SS-V0 and $\tilde{t} > 15000$ for SS-EXP), the asymptotic self-similar decay regime is entered, in Figure \ref{fi:selfsim} we show several azimuthal velocity profiles, normalized by their mean value. For the SS-V0 case, the normalized velocity profiles at times $\tilde{t}=5000$ to the end of our simulation $\tilde{t}=7500$ (corresponding to $t\nu/d^2\approx 0.2$) collapse, and agree with the theoretically calculated fundamental Bessel function, while the earlier normalized profiles between $\tilde{t}\approx1000$ and $\tilde{t}\approx5000$ show slight deviations as the asymptotic stage has not yet been reached. The normalized profile around $\tilde{t}\approx500$ is very different, as it is in the non-normal transient growth state. For the SS-EXP case, the asymptotic, self-similar stage is only reached for $\tilde{t}\approx 15000$, corresponding to $t\nu/d^2 \approx 0.25$.

When comparing these profiles to the proposed self-similarity in Ref.~\cite{ver16}, we find that  they are similar at \emph{intermediate} times: a flat profile in the center, indicating strong turbulent mixing, and a steeper boundary layer at the outer cylinder, due to the centrifugal instability at the outer cylinder and the centrifigual stability at the inner cylinder. This intuition coincides with the fact that the largest reported profile at $t=100s$ is only $t\nu/d^2 \approx 0.03$ in viscous units. The proposed self-similarity is shown for a small interval of time ($0.001< t\nu/d^2 < 0.03$), and may be just be a product of the relatively small time interval.

In summary, in this manuscript, we have studied the decay of Taylor-Couette turbulence, focusing also on the early and late stages of decay that could not be observed in Ref. \cite{ver16}. We have shown that three distinct life stages are seen where different decay mechanisms dominate. In the first life stage, the energy decays mainly through the linearly unstable modes, i.e.~the rolls. After this, the available energy is directed towards non-normal modes which sustain transient growth. During the transition between stages, a redistribution of energy between structures occurs, and $\langle u_\theta \rangle^2_\Omega$ \emph{increases}. While this can only be seen if the flow inertia is small, it highlights that the decay process is anisotropic and be temporally non-monotonic. Finally, the decay becomes purely viscous, in spite of the relatively high instantaneous Reynolds number, and self-similarity can be observed. We have shown that no simple power law can cover all three stages. The most unstable modes dominate the early decay; the modes decay progressively, from linearly-unstable modes with roll-like instabilities, to non-normal transient growth to self-similar decay, which behaves like a quenching problem.

This progression of decaying modes from more to less unstable can shed light onto how real-world unforced turbulence decays. Our finding that anisotropic flows can decay in a viscid way even at moderate Reynolds numbers, that redistribution of energy between structures can lead to non-monotonic behavior, and  that wall-friction is a secondary mechanism in the initial decay can provide insight into the decay of geo- and astrophysically relevant systems even if confinement of the Taylor-Couette system could still play a role. Further insights into decay and the cross-over between the first two stages can be provided by studying the decay in linearly stable Taylor-Couette flow \cite{ost16b}, or in plane Couette flow, i.e. the flow between two parallel plates, which could make the first stage of decay less important. The TC geometries studied have very limited curvature, and its (de)stabilizing role in reducing the non-normal transient growth can be explored. Finally, studies of the decay of thermal turbulence \cite{ahl09}, and further exploring the analogy between TC and Rayleigh-B\'enard convection \cite{dub02,eck07b}, the flow in a layer heated from below and cooled from above, which have been called the ``twins of turblence research'' \cite{bus12}, is another research line which can lead to increased understanding of decaying geo- and astrophysical turbulence. 

\begin{acknowledgments}

\emph{Acknowledgments:} We thank B. Eckhardt for the fruitful and stimulating discussions, and R. A. Verschoof for providing the data from Ref.~\cite{ver16}. This research was  supported in part by the National Science Foundation under Grant No. PHY11-25915. We also gratefully acknowledge computational time for the simulations provided by SURFsara on resource Cartesius through a NWO grant. 
\end{acknowledgments}

\bibliography{literatur}

\end{document}


\title{Three life stages of wall-bounded decay of Taylor-Couette turbulence: reanalysis of the experimental data}

\author{Rodolfo Ostilla-M\'{o}nico}\email{rostillamonico@g.harvard.edu}
\affiliation{School of Engineering and Applied Sciences, Harvard University, Cambridge, MA 02138, USA.}
             
\author{Xiaojue Zhu}
\affiliation{Physics of Fluids Group, Faculty of Science and Technology, MESA+ Research
             Institute, and J. M. Burgers Centre for Fluid Dynamics,
             University of Twente, PO Box 217, 7500 AE Enschede, The Netherlands.}

\author{Vamsi Spandan}
\affiliation{Physics of Fluids Group, Faculty of Science and Technology, MESA+ Research
             Institute, and J. M. Burgers Centre for Fluid Dynamics,
             University of Twente, PO Box 217, 7500 AE Enschede, The Netherlands.}

\author{Roberto Verzicco}
\affiliation{Dipartimento di Ingegneria Industriale, University of Rome ``Tor Vergata", Via del Politecnico 1, Roma 00133, Italy}
\affiliation{Physics of Fluids Group, Faculty of Science and Technology, MESA+ Research
             Institute, and J. M. Burgers Centre for Fluid Dynamics,
             University of Twente, PO Box 217, 7500 AE Enschede, The Netherlands.}

\author{Detlef Lohse}
\affiliation{Physics of Fluids Group, Faculty of Science and Technology, MESA+ Research
             Institute, and J. M. Burgers Centre for Fluid Dynamics,
             University of Twente, PO Box 217, 7500 AE Enschede, The Netherlands.}
\affiliation{Max Planck Institute for Dynamics and Self-Organisation, 37077 G\"ottingen, Germany. }

\date{\today}

\begin{abstract}
In this addendum, we detail a reanalysis of the experimental data from Verschoof \emph{et al.} \cite{ver16}, and show that for most of the duration, it corresponds to the same self-similar regime seen in our simulations, with a particular focus on the dissipation and Reynolds numbers.
\end{abstract}

\pacs{47.27.nf, 47.32.Ef}

\maketitle

We already mentioned in the main manuscript that there are two main differences, aside from the Reynolds number and other geometrical parameters, between our simulation and the experiment of Ref. \cite{ver16}. First, we instantaneously stop the cylinders, and second,  in the simulation we set the velocity of both cylinders to the mean velocity, i.e. zero in the rotating frame, while in the experiment the inner cylinder is brought to the same velocity as the outer cylinder, and there is a substantial mean momentum which must be transported from bulk to cylinders. 

We postulate in the manuscript that most of the physics is hidden as the stopping takes too long, and that in Ref. \cite{ver16} most of what is seen simply corresponds to the late life-stage of self-similar viscous decay. In this addendum we show

We find that the self-similar regime observed and discussed in Ref.~\cite{ver16} corresponds to a \emph{late} stage of decay, which can be described as a quenching problem. Prior to this viscosity-dominated stage, two earlier life stages of decay are observed, which are dominated by linear instabilities and non-normal transient growth, respectively.

The simulations were run until a statistically stationary state was reached, where two roll pairs could be seen in the velocity fields, and the torque at both cylinders was equal when temporally averaging over a sufficiently long time. For the first case, which we denote as sudden stop (SS), we instantaneously stop the cylinders at $t=0$, and  allow the turbulence to decay. Two differences exist between the simulation and the experiment. First, we stop the cylinders instantaneously, while the stopping time of the experiment by Verschoof \emph{et al.}~\cite{ver16} is around 12 s. Even if this is still orders of magnitude smaller than the viscous time ($d^2/\nu=6\cdot 10^3$ s) it correspo

To eliminate as far as possible the effects of the wall, we also run a second case, denoted by sudden disengagement (SD) in which we suddenly change the boundary condition at the wall to stress-free. The flow is no longer forced by shear, and thus the turbulence decays. This boundary condition prevents the system from decaying to a stationary state of rigid rotation, because for rigid rotation there is still a shear stress at the walls.

As turbulence decays, the computational method is switched for efficiency. At $\tilde{t} = tU/d \approx 125$ for the no-slip case, and $\tilde{t} \approx 325$ for the free-slip case, the flow fields were down-sampled to a resolution of $512^3$, and the treatment of the viscous terms in the homogeneous directions was made implicit. This allowed the time-step to increase to $\Delta \tilde{t} = 1$, dramatically reducing the computational cost. Both simulations were then advanced in time up to $\tilde{t}=7500$, late enough to be in the asymptotic stage.


\begin{figure}
\includegraphics[width=0.48\textwidth]{pdffigs/1.eps}
\centering
\caption{Top: Temporal evolution of the kinetic energy in the azimuthal velocity (red) and the wind velocity (blue) for the SD (solid) and SS (dashed) cases. Bottom: Temporal evolution of the energy dissipation. The figures show the same data in semilogarithmic scale (left) and double logarithmic scale (right). The inset on the top-right figure shows a zoom-in of the early-stage decay. }
\label{fi:urms} 
\end{figure}

\begin{figure*}
\centering
\includegraphics[width=0.6\textwidth]{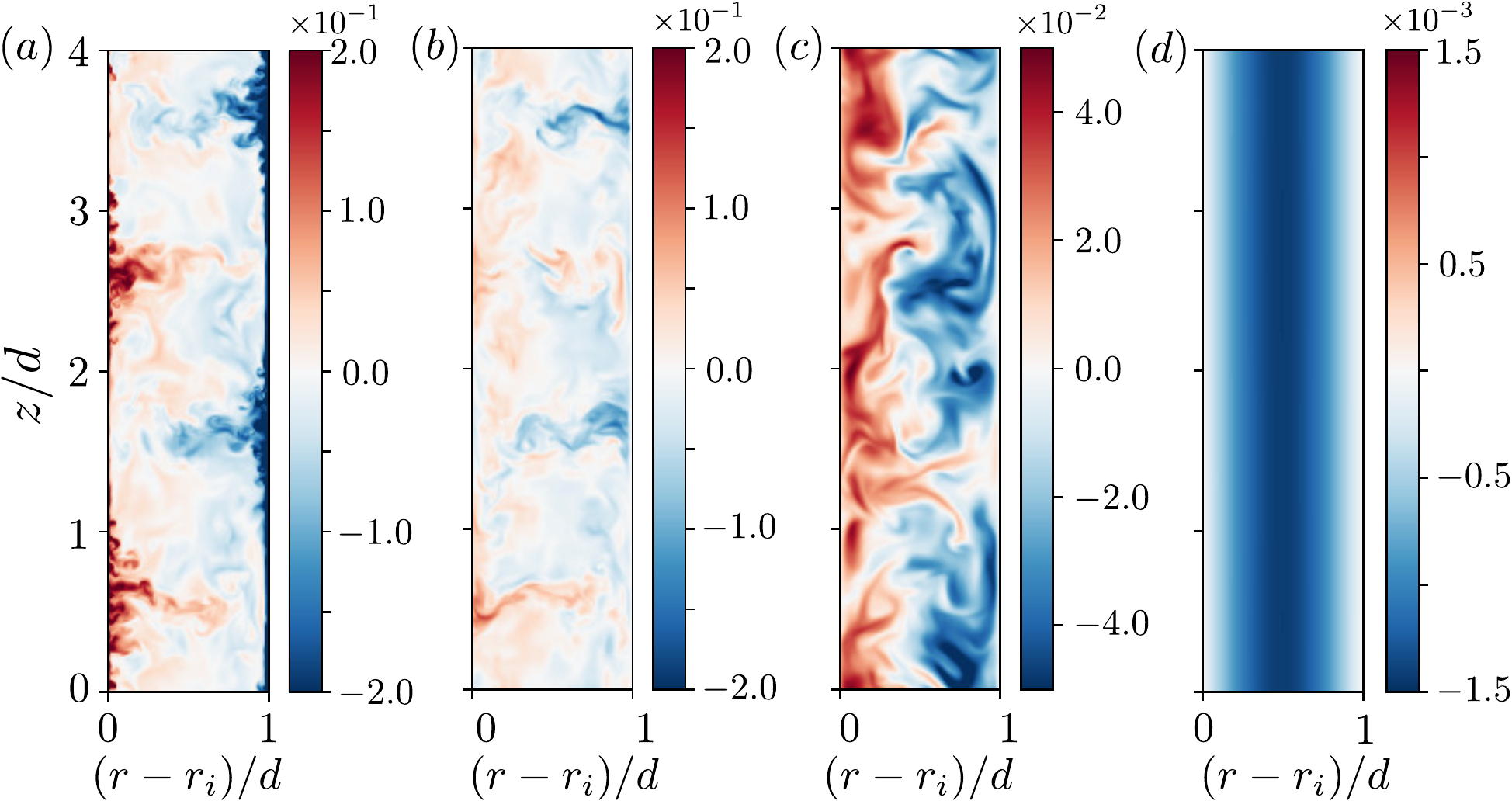}
\caption{Pseudocolor visualization of the azimuthal velocity in a constant-azimuth cut for the SS case at four different times (left to right): $\tilde{t}=0$, $10$, $75$ and $5000$. }
\label{fi:vizthcut} 
\end{figure*}

Figure \ref{fi:urms} shows the temporal evolution of the average kinetic energy of the azimuthal velocity ($\frac{1}{2}\langle u_\theta^2\rangle_\Omega$) and the wind kinetic energy ($\frac{1}{2}\langle u_r^2 + u_z^2 \rangle_\Omega$) for both the SS and the SD cases, and also the temporal evolution of the total kinetic energy dissipation rate $\epsilon=\frac{1}{2}\nu\langle \partial_i u_{j}\rangle^2_\Omega$, where $\langle \dots \rangle$ denotes the averaging operator and $\Omega$ the entire fluid volume.  No overarching behavior or power law which describes either the dissipation or the evolution of the kinetic energy can be seen. Instead, three different life stages of decay are revealed, which we detail here. These stages are illustrated by Figure \ref{fi:vizthcut}, which shows a visualization of the azimuthal velocity at four different time instants for the SS case, from the start of the simulation to the asymptotic state dominated by viscous diffusion. Aside from the magnitude of the velocity being different, we can also see wide variation in the flow topology. A similar topology is seen for the SD case, too.

During the first life stage, which takes place between $\tilde{t}=0$ to $\tilde{t}\approx10$ (corresponding to Figure \ref{fi:vizthcut}\emph{b}), the large-scale rolls remain in motion, as they are still being fed from the perturbations inside the boundary layers. The inset of Figure \ref{fi:urms} shows that the wind kinetic energy remains constant (horizontal) between $\tilde{t}=0$ and $\tilde{t}\approx 10$, while the kinetic energy of the azimuthal velocity decreases by more than an order of magnitude. The different behavior of velocity components highlights the anisotropic character of the decay.

Further proof of this is seen in figure \ref{fi:kinen_s1}a, which shows the distribution of the azimuthal kinetic energy at the start and end of the first stage. At the start of the decay, energy is concentrated near the boundary layers. Once this energy is drained by the rolls, the rolls fade away. The boundary layers grow, and the fluctuation maximum moves away from the walls. Energy is redistributed between the existing structures, and the flow topology changes from Figure \ref{fi:vizthcut}b to \ref{fi:vizthcut}c. Counterintuitively, the decay of azimuthal kinetic energy is non-monotonic, and we see an \emph{increase} in azimuthal kinetic energy at around $\tilde{t}\approx 15$ which is drawn from the wind kinetic energy during the redistribution. 

Wall-friction plays a secondary role during this stage. The main player draining the available energy in the boundary layer are the rolls. The wall-friction comes in as a small correction, making the SS case undergo a transition from the first life stage to the second life stage slightly faster than the SD case. Non-monotonic behavior can be seen for both SS and SD cases.

\begin{figure}
\centering
\includegraphics[width=0.48\textwidth]{pdffigs/3.eps}
\caption{Left: Azimuthally and axially averaged azimuthal kinetic energy dissipation for the SS case, and $\tilde{t}=0$ (blue), $\tilde{t}=10$ (orange), $\tilde{t}=30$ (green) and $\tilde{t}=50$ (red). Right: Axially and azimuthally averaged normalized azimuthal velocity profiles for the SS at $\tilde{t}=500$ (blue), $\tilde{t}=1000$ (green), $\tilde{t}=5000$ (red circles) and $\tilde{t}=7500$ (orange), as well as the theoretical solution based on Bessel functions (black dashed). }
\label{fi:kinen_s1} 
\end{figure}

Then the second life stage is entered which takes place between $\tilde{t}\approx 10 $ and $\tilde{t} \approx 500$. With the rolls faded away, perturbations are amplified non-normally. A detailed discussion of this non-normal-non-linear mechanism is found in Ref. \cite{gro00rmp}. Here it is sufficient to state that perturbations in the axial (spanwise) direction tend to grow faster than those in other directions. As typical for this type of instability, the growth of these perturbations is transient, and after several time units they decay. For the SS case, the wind kinetic energy drops to a minimum around $\langle u^2 \rangle_\Omega \approx 10^{-5}$ and it remains at that level until the end of this regime, while for the SD case, the wind kinetic energy does not reach a plateau but instead oscillations appear throughout the entire decay. The SS case appears to drain the available energy of the many non-normal modes in a shorter amount of time, while this draining occurs over a longer time-scale for the SD case. The signature of non-normal transient decay can be clearly seen for the wind kinetic energy in the SD case. When one of the non-normal modes grows, amplifies transiently and decays, it imprints the oscillations seen in the wind kinetic energy for $\tilde{t}>100$.

\begin{figure}
\centering
\includegraphics[width=0.4\textwidth]{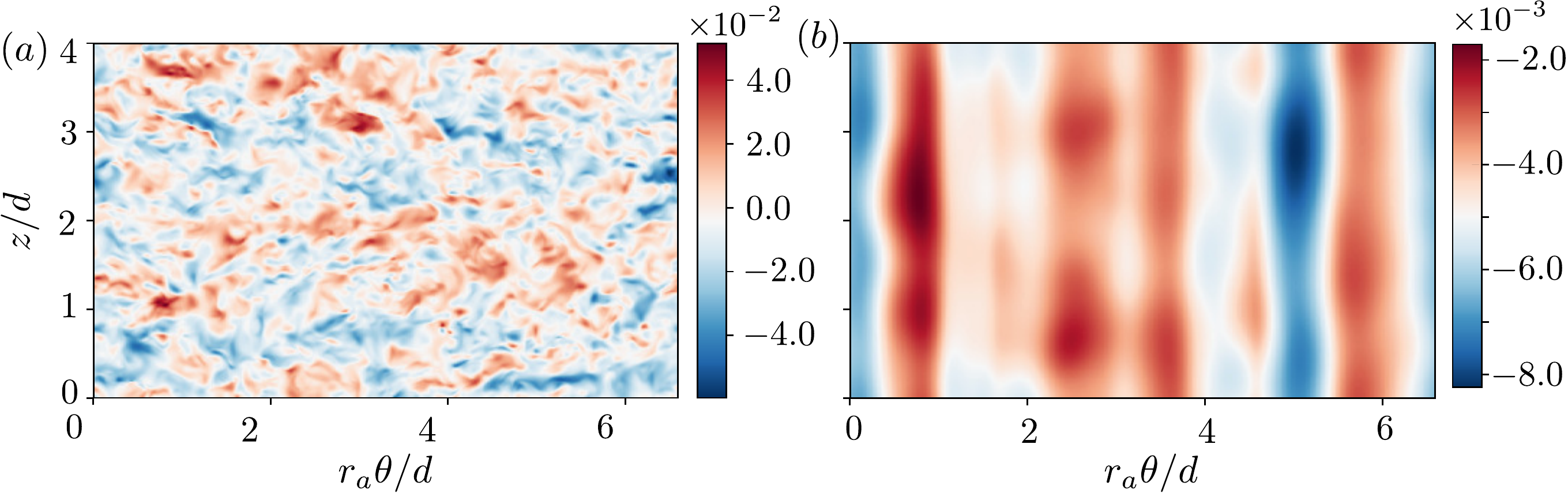}
\caption{Pseudocolor visualization of the azimuthal velocity in a constant-radius cut at the mid-gap ($r=r_a$) for the SS case at two different times $\tilde{t}=75$ (top) and $1000$ (bottom). }
\label{fi:vizrcut} 
\end{figure}

Due to the non-normal transient growth and dissipation of axially-oriented perturbations, the flow becomes practically homogeneous in the axial direction for $\tilde{t} > 500$. The large-scale flow structure is now columnar. This is shown for the SS case in Figure \ref{fi:vizrcut}, and the same happens for the SD case (not shown). The role of wall-friction is minor in the transition between the second and third stage, as this happens when transient non-normal growth is exhausted. The azimuthal direction also becomes practically homogeneous around $\tilde{t} > 2000$. Then the decay enters its last life stage, in which the decay is dominated by viscosity. 

In this last life stage the energy dissipation rate behaves in a quasi-exponential manner. The wind kinetic energy is negligible and we can expect the azimuthal velocity to behave like a passive scalar with diffusion. This is analogous to the quenching problem of a cylindrical annulus, whose solution is given by $u_\theta(r,t)=\sum_n A_n J_0(\lambda_n r)\exp(-\nu \lambda_n^2 t)$, where $J_0(r)$ are Bessel functions of the zeroth order and $\lambda_n$ their associated eigenvalues, and $A_n$ depend on the initial conditions \cite{prosbook}. If $t$ is large, only the slowest-decaying function with the smallest eigenvalue has significant energy. The energy dissipation behaves exponentially in time, and self-similarity comes out naturally. To demonstrate that beyond $\tilde{t} > 5000$, the asymptotic self-similar decay regime is entered, in Figure \ref{fi:kinen_s1}b we show several azimuthal velocity profiles for the SS case, normalized by their mean value. The normalized velocity profiles at times $\tilde{t}=5000$ to the end of our simulation $\tilde{t}=7500$ (corresponding to $t\nu/d^2\approx 0.2$) collapse, and agree with the theoretically calculated fundamental Bessel function, while the earlier normalized profiles between $\tilde{t}\approx1000$ and $\tilde{t}\approx5000$ show slight deviations as the asymptotic stage has not yet been reached. The normalized profile around $\tilde{t}\approx500$ is very different, as it is in the non-normal transient growth state.

Wall friction becomes dominant only once the mechanisms for non-normal transient growth are exhausted. For the above simulations, this corresponds to instantaneous Reynolds numbers of about $Re\sim\mathcal{O}(10^2)$. This Reynolds number could potentially be higher for other turbulent flows. For example, in the experiment of Ref.~\cite{ver16}, the self-similarity seen in the velocity profiles suggests that the decay is in the last life stage, even when the instantaneous Reynolds numbers is still quite high.

In summary, in this manuscript, we have studied the decay of Taylor-Couette turbulence, focusing also on the early stage decay that could not be observed in Ref. \cite{ver16}. We have shown that three distinct life stages are seen where different decay mechanisms dominate. In the first life stage, the energy decays mainly through the linearly unstable modes, i.e.~the rolls. After this, the available energy is directed towards non-normal modes which sustain transient growth. During the transition between stages, a redistribution of energy between structures occurs, and $\langle u_\theta \rangle^2_\Omega$ \emph{increases}, highlighting that the decay process is highly anisotropic and can incur in temporal non-monotonicity. Finally, the decay becomes purely viscous, in spite of the relatively high instantaneous Reynolds number, and self-similarity can be observed. We have shown that no simple power law can cover all three stages. The most unstable modes dominate the early decay; the modes decay progressively, from linearly-unstable modes with roll-like instabilities, to non-normal transient growth to self-similar decay, which behaves like a quenching problem. 

This progression of decaying modes from more to less unstable can shed light onto how real-world unforced turbulence decays. Our finding that anisotropic flows can decay in a viscid way even at moderate Reynolds numbers, that redistribution of energy between structures can lead to non-monotonic behavior, and  that wall-friction is a secondary mechanism in the initial decay can provide insight into the decay of geo- and astrophysically relevant systems. Further insights into decay and the cross-over between the first two stages can be provided by studying the decay in linearly stable Taylor-Couette flow \cite{ost16b}, or in plane Couette flow, i.e. the flow between two parallel plates, which could make the first stage of decay less important. The TC geometries studied have very limited curvature, and its (de)stabilizing role in reducing the non-normal transient growth can be explored. Finally, studies of the decay of thermal turbulence \cite{ahl09}, and further exploring the analogy between TC and Rayleigh-B\'enard convection \cite{dub02,eck07b}, the flow in a layer heated from below and cooled from above, which have been called the ``twins of turblence research'' \cite{bus12}, is another research line which can lead to increased understanding of decaying geo- and astrophysical turbulence. 

\begin{acknowledgments}

\emph{Acknowledgments:} We thank B. Eckhardt for the fruitful and stimulating discussions. This research was  supported in part by the National Science Foundation under Grant No. PHY11-25915. We also gratefully acknowledge computational time for the simulations provided by SurfSARA on resource Cartesius through a NWO grant. 
\end{acknowledgments}

\bibliography{literatur}